%% file: ms.tex
\documentclass[12pt,preprint]{aastex}
\usepackage{graphicx}

\newcommand{\invday}{\,\rm{d}^{-1}}
\newcommand{\porb}{P_{\rm{orb}}}

\begin{document}

\title{Ellipsoidal Variability in the OGLE Planetary Transit Candidates}

\author{E. Sirko \& B. Paczy\'nski}
\affil{Princeton University Observatory, Princeton, NJ 08544--1001, USA}
\affil{E-mail: esirko@astro.princeton.edu}
\affil{E-mail: bp@astro.princeton.edu}

\begin{abstract}
We analyze the photometry of 117 OGLE stars with periodic transit events
for the presence of ellipsoidal light variations, which indicate the presence
of massive companions.  We find that $ \sim 50\% $ of objects may
have stellar companions, mostly among the short period systems.  In our
Table 1 we identify a coefficient of ellipsoidal variability for
each star, $a_{c2}$, which can be used to select
prime candidates for planetary searches.  There is a
prospect of improving the analysis, and the systems with smaller ellipsoidal
variability will be identified, when the correlations in the OGLE photometry
are corrected for in the future, thereby providing a cleaner list of systems
with possible planets.
\end{abstract}


\section{Introduction}

The first extra solar planet discovered to exhibit photometric transits was
HD 209458 (Charbonneau et al. 2000, Henry et al. 2000), but its orbit had 
been first determined spectroscopically (Mazeh et al. 2000).  Massive efforts
to detect photometric transits on their own put at least 23 teams into
the competition (Horne 2003).  By far the largest list of periodic transit
candidates was published by the OGLE team (Udalski et al. 2002a,b,c), with
a total of 121 objects, with all photometric data available on the Web.
The first confirmation that at least one of these has a `hot
Jupiter' planet was obtained by Konacki et al. (2003).  They also found that
a large fraction of OGLE candidates were ordinary eclipsing binaries
blended with a brighter star which was not variable, and its light diluted
the depth of the eclipses, so they appeared as shallow transits.

It was clear from the beginning that many transits may be due to red dwarfs
or brown dwarfs, and that some of these massive companions may give rise to
ellipsoidal light variability (Udalski et al. 2002a).  In fact in that
paper OGLE-TR-5 and OGLE-TR-16 were noted to exhibit such a variability,
indicating that 
the companion mass had to be substantial.  Ellipsoidal variability
is a well known phenomenon among binary stars (cf. Shobbrook et al. 1969,
and references therein).  Tidal effects are responsible for making a star
elongated toward the companion.  Ellipsoidal variability is due to the
changes in the angular size of the distorted star and to gravity 
darkening, with half of the orbital period.  Obviously,
the more massive the companion, and the closer it is to the primary,
the stronger the
effect.  It can be used to remove from the OGLE sample objects which have
massive, and therefore not planetary, companions (Drake 2003).

The aim of this paper is to extend the work of Drake (2003) to the new
list of OGLE transit candidates (Udalski 2002c) and to provide a more
realistic error analysis, which is needed in order to assess the reality
of ellipsoidal variability.

\section{Data Analysis}

We took all photometric data from the OGLE Web site:

\centerline{http://sirius.astrouw.edu.pl/\~{}ogle/}

\centerline{http://bulge.princeton.edu/\~{}ogle/}

\noindent
There are three data sets: two provide a total of 59 objects in the field
close to the Galactic Center (Udalski et al. 2002a,b), and the third
provides 62 objects in a field in the Galactic Disk in Carina (Udalski
et al. 2002c).  There are typically 800 data points for OGLE-TR-1 -
OGLE-TR-59, and 1,150 data points for OGLE-TR-60 - OGLE-TR-121.  The 
Carina data set had somewhat longer exposures and the field was much
less crowded, so the photometric accuracy is somewhat better than in
the Galactic Center field.

To analyze the light curves for ellipsoidal variability the data points
within the transits must first be removed.  Our algorithm was to 
start at mid-transit, working outwards, and to reject data points 
until the third data point
for which the baseline magnitude was within its photometric error bars.
We then obtained a five parameter fit to the rest of the data:
$$
I_k = <I> + a_{c1} \cos p_k + a_{s1} \sin p_k 
+ a_{c2} \cos 2 p_k + a_{s2} \sin 2 p_k ,
\eqno(1)
$$
where $ I_k $ is data point number $ k $, and $ p_k $ is its
phase calculated with the orbital period provided by Udalski et al. 
(2002a,b,c).  The phase is $ p = 0 $ at mid transit.  The values of all five
parameters: $ <I> $, $ a_{c1} $, $ a_{s1} $, $ a_{c2} $, and $ a_{s2} $
were calculated with a least squares method.  The formal errors for all
the sinusoidal coefficients for a given star were practically the same.  
Fig. 1 shows period-folded light curves of several objects, with the 
best-fit five-parameter function (eq. 1) over plotted.  Ellipsoidal
variability is clearly visible in objects such as OGLE-TR-16 as a
sinusoidal component with half the period of the binary and with 
the phase of minimum light flux being the same as the phase of the transit.

In the five parameter fit given with eq. (1) the term $ a_{c2} $ 
corresponds to the ellipsoidal light variations.  Tidal effects make
a star the brightest at phases 0.25 and 0.75, and the dimmest at phases
0.0 and 0.5.  This corresponds to $ a_{c2} > 0 $, as the $ I $ magnitude is
largest when the star is at its dimmest.  The values of $ a_{c2} $ with their
nominal error bars are shown in Fig. 2.  Note that for a number of stars
the coefficient is negative, many standard deviations smaller than zero.
This clearly shows that the nominal errors are not realistic.  This is
understandable, as there are strong correlations between
consecutive photometric  data points (Kruszewski \& Semeniuk 2003).

The rms deviation between
the five parameter fit and the data is a measure of the photometric
errors for individual measurements.  The rms errors and the average of 
the formal errors 
of the four sinusoidal coefficients of the fit are shown in Fig. 3 
with circles and triangles, respectively.  
The formal errors for the fitted parameters are approximately 
$ N^{1/2} $ times smaller than the rms values, as expected;
$ N $ is the number of photometric data points for a given star.
Open symbols refer to the stars in the Galactic Center field, and filled
symbols to the stars in Carina.  The crosses are error estimates 
by Drake (2003, Table 1) for stars in the Galactic Center field.
The squares will be explained later.

It is clear that the errors increase with the $ I $ magnitude, as expected.
The errors are smaller for the Carina stars, as expected.  Two stars,
OGLE-TR-24 and OGLE-TR-58, show anomalously large rms.  Inspection of their
light curves shows they exhibit long term changes of about 0.015
magnitudes.  The nature of this variability is not known, and we exclude
these two objects from the error histogram shown in Fig. 5, discussed below.
The orbital periods are not known for OGLE transits 43, 44, 45,
and 46, so these stars are not considered in this paper.
The following analysis was done for the remaining 117 objects. 
Note that objects 8 and 29 are the same star, but they are treated
separately in the OGLE database, so we treat them separately here as well. 

The planetary transit candidates have orbital periods in the range 
$ 0.57 - 9.2 $
days, which corresponds to the frequency of ellipsoidal variability in
the range $ 0.22 - 3.5 \invday $.  We calculated power spectra for every
star as follows:
$$
p_{\nu} = a_{c, \nu }^2 + a_{s, \nu}^2 , \hskip 1.0cm
\nu = { i \over 250 ~ {\rm day } } , \hskip 1.0cm
i = 1, 2, 3, .... 1000,
\eqno(2)
$$
where
$$
a_{c, \nu } = 
~ { \sum _{k=1}^N (I_k-<I>) \cos \left( 2 \pi \nu t_k \right) \over
    \sum _{k=1}^N  \cos ^2 \left( 2 \pi \nu t_k \right) } ,
\hskip 1.0cm
a_{s, \nu } = 
 ~ { \sum _{k=1}^N (I_k-<I>) \sin \left( 2 \pi \nu t_k \right) \over
     \sum _{k=1}^N  \sin ^2 \left( 2 \pi \nu t_k \right) } ,
\eqno(3)
$$
where the summation
is done over all photometric data points for a given star, excluding the
data points within the transits.

An example of the power spectrum is shown in Fig. 4 for OGLE-TR-5.
Also shown is a power law fit to the spectrum:
$$
p_{ \nu ,f} = p_0 ~ \nu ^{p_1} ,
\eqno(4)
$$
where the two parameters $ p_0 $ and $ p_1 $ were calculated for each
star using a least squares method.  The binary period for OGLE-TR-5 is given
by Udalski et al. (2002a) as 0.8082 days.  The arrows 
correspond to the orbital frequency $ 1 / \porb = 1.237 \invday $,
and the expected frequency of ellipsoidal variability
$ 2 / \porb = 2.475 \invday $.

We calculated power spectra for all stars, and each was fitted with its 
own power law.  For every star and for every frequency we calculated
the coefficients $ a_{c, \nu } $ and $ a_{s, \nu } $, as defined
by eq.~(3), and divided each by
$ b_{ \nu} \equiv ( p_{ \nu , f} / 2 )^{1/2} $ to normalize it.
A histogram of these normalized terms is shown in Fig. 5.
Also shown is Gaussian distribution with unit variance; it is
fairly similar to the histogram, which implies that $ b_{ \nu } $
provides a reasonable estimate of the statistical error.

Our power spectrum formula (eqs. 2 and 3) can be derived by a least squares
method at each frequency.  It is similar to the Lomb-Scargle (LS) 
periodogram formula (Lomb 1976, Scargle 1982, Press et al. 1992), but 
differs in several respects.  Perhaps most importantly for this work,
the LS periodogram does not distinguish between the sine and cosine terms.
Therefore, a phase offset $\tau$ is necessary in the LS periodogram
formula, but not in our power spectrum since we fit $a_{c,\nu}$ and 
$a_{s,\nu}$ separately.  One strength of the LS periodogram is a prescription
to determine the significance level of any peak in the power spectrum.
However, this prescription is most useful when the resonant frequencies
are unknown beforehand, and requires a knowledge of the number of
effectively independent frequencies $M$ in the power spectrum, because 
as $M$ increases, the probability that spurious noise looks like 
genuine signals increases.  However, in our case, we know the expected
frequency of ellipsoidal variability beforehand, so the value of $M$ is
irrelevant for our simple analysis.  As shown in Fig.~5,
we are able to define an error $b_{\nu}$ for each value of $a_{c,\nu}$
and $a_{s,\nu}$ which is approximately Gaussian, so that the usual
(68, 95, 99.7)\%, etc. confidence values `approximately' apply for values of 
$a_{c,\nu}$ and $a_{s,\nu}$ within (1, 2, 3)$\sigma$, etc.  In other words,
a value of $a_{c2}$ that is $3\sigma$ away from zero indicates
ellipsoidal variability at a 99.7\% confidence level, depending on how
much faith one puts into the Gaussian nature of Fig. 5.  Note that
the tails of the distribution are exaggerated on the logarithmic axis,
and that some contribution to the tails should be from genuine signal.

For every star we calculated the $ a_{c1} $ and $ a_{c2} $ terms of eq. (1)
and estimated their errors $b_1$ and $b_2$ as $b_{ \nu } $ evaluated at the
corresponding frequencies.  The errors for the $a_{c2}$ term, $b_2$,
are shown in Fig. 3 as squares.
It is clear that these statistical errors are much larger than the
errors shown as triangles, which were based on the assumption that 
all photometric data points are uncorrelated.  The `triangle' error
bars were used in our Fig. 2, which provided the first hint that they are
underestimates of the true errors, as so many values of the ellipsoidal
variability parameter $ a_{c2} $ were negative with very small `triangle'
error bars.

The two amplitudes $ a_{c2} $ and $ a_{c1} $ are shown in Fig. 6 and Fig. 7,
respectively, as a function of orbital period.  Negative values of $ a_{c2} $
are physically meaningless, and presumably these are $ a_{c2} \approx 0 $
which were `scattered' to negative values by statistical errors.  Indeed,
there are 29 negative values, with 20 within one $ \sigma $ of zero, and
9 outside of one $ \sigma $, with the ratio 20/9 close to that
expected for a Gaussian distribution.
This also implies that our error estimate is realistic; here and in 
the following we take $\sigma = b_1$ for $a_{c1}$ and $a_{s1}$, 
$\sigma = b_2$ for $a_{c2}$ and $a_{s2}$.

We expect that about the same number, 29, of positive $ a_{c2} $ values 
is due to errors, while the remaining $ 117 - 29 - 29 = 59 $ are real
ellipsoidal variables.  This estimate suggests that $ \sim 50\% $ of all
OGLE transit candidates have massive companions, not planets.
Tidal effects responsible for ellipsoidal
variations increase strongly with reduced orbit size, and therefore they
are expected to be more common at short orbital periods.  Indeed, this
appears to be the case in Fig. 6.

Ellipsoidal variability scales with the mass ratio, and for low mass
companions like planets it becomes much smaller than our errors.  Hence,
all OGLE transit stars with measurable ellipsoidal variability should
be excluded from the list of planetary candidates.  Note that blending
with non-variable stars dilutes the variable component, and may suppress
the amplitude of the apparent ellipsoidal variability.

Some stars vary with the orbital period.  Positive values of $ a_{c1} $, 
as shown in Fig. 7, may indicate a heating (reflection)
effect of the companion by the primary, as in the well known case of Algol,
or they may indicate that the true orbital period is twice longer than
listed by OGLE, and the variability is ellipsoidal.  In the former case
nothing can be said about the companion mass; in the latter case the companion
is too massive to be a planet.  

Note that there are 44 stars in Fig. 7 with negative values of $ a_{c1} $,
and 27 of these are within their $ 1 \sigma $ error bars of zero.  This
is consistent with no credible negative $ a_{c1} $.  As the errors are
expected to be symmetric there must be $ \sim 44 $ stars which nominally
have positive values of $ a_{c1} $, but in fact are consistent with
$ a_{c1} = 0 $.  Unfortunately,
we cannot point to the remaining 29 stars which may have real
`reflection' effects, except for OGLE-TR-39,
which has the orbital period given as 0.8 days, and it obviously has
$ a_{c1} > 0 $.  This star shows large values 
(in terms of their errors)
of both terms: $ a_{c1} $ and $ a_{c2} $.  This indicates that the
orbital period is correct, the companion has its hemisphere facing the
primary noticeably heated, and its mass is large enough to induce ellipsoidal
light variations.  The reflection effect is expected to be stronger
for binaries with small separations, i.e. short orbital period.  There
is some evidence for this effect in Fig. 7.

Table 1 lists all OGLE transit candidates with known orbital periods,
giving their OGLE-TR number, average magnitude $ <I> $, orbital period
$ \porb $ in days, the $ a_{c1} $ and $ a_{c2} $ coefficients in 
milli-magnitudes, and
the number of transits detected by OGLE, $N_{\rm{tr}}$. 
Objects with few detected transits are more likely to have
incorrect periods.  Therefore, the ellipsoidal variability of the objects 
with $N_{\rm{tr}} \sim 2$ is more likely to go undetected in our
analysis (but possibly could be detected in the $a_{c1}$ term if
the incorrect period happened to be half the true period).
The $ a_{c1} $ and $ a_{c2} $ coefficients are expressed in terms of their
errors in Table 1, in the format $b_i(a_{ci}/b_i \pm 1)$,
so that objects with larger values of $a_{c2}/b_2$ are more
likely to be ellipsoidal variable, and those with smaller or negative 
values are better planetary system candidates.
It should become possible to recognize more objects with bona fide 
ellipsoidal light variations when the systematic errors are reduced
by a more thorough analysis carried out by Kruszewski \& Semeniuk (2003).

It is interesting to plot $ a_{c1} $ versus $ a_{c2} $, as shown in Fig. 8.
It is clear that while there are several possibly real positive $ a_{c1} $
terms, there are considerably more positive $ a_{c2} $ terms.

In a simple model where the presence of a companion may give rise to
ellipsoidal light variations and the `reflection' effect the terms
$ a_{s1} $ and $ a_{s2} $ in eq. (1) should be zero.  Fig. 9 presents
these coefficients in units of their errors.  There are 29 stars with
$ a_{s2} $ having absolute value larger than one $ \sigma $, with
88 values smaller than one $ \sigma $.  The corresponding numbers for
$ a_{s1} $ are 41 and 76, respectively.  This is close to the ratio
expected if the true values of both coefficients were zero, and their
errors were Gaussian.  The average and the rms values are:
$ < a_{s1} /b_1 >\ = -0.27 $, $ < a_{s2} / b_2 >\ = +0.10 $,
$ <(a_{s1}/b_1)^2>^{1/2}\ = 1.17 $, and $ <(a_{s2}/b_2)^2>^{1/2}\ = 0.93 $.
All this implies that our error estimate is reasonable.  Note that
OGLE-TR-68, with $ a_{s1}/b_1 = -4.4 $, may have a real variability,
possibly induced by a spot on the star.  

By treating the duplicate objects OGLE-TR-8 and OGLE-TR-29 separately
in this analysis, we are able to verify the consistency of the algorithm.
As can be seen from Table 1, for these two objects 
the cosine coefficients $a_{c1}$ and
$a_{c2}$ are consistent within error.  Furthermore, for OGLE-TR-8, 
$a_{s1} = -0.006 \pm 1.47, a_{s2} = 0.643 \pm 1.32$, and for OGLE-TR-29,
$a_{s1} = -0.550 \pm 1.71, a_{s2} = 0.664 \pm 1.55$, so the sine 
coefficients are also consistent within error.

\section{Discussion}

It is interesting to compare our errors, given in Table 1, 
with those estimated by Drake (2003, Table 1, Galactic Center region only).  
Our errors are also shown in Fig. 3 as squares, while Drake's errors are 
shown with crosses.  For the faintest stars
systematic errors are comparable to the photon noise, hence there is
only a small difference, approximately a factor of 2.  For the bright stars
photon noise is negligible compared to systematic errors, and our estimate 
is up to 5 times larger than Drake's (OGLE-TR-16).  We stress that our 
estimate is realistic, as demonstrated with Fig.~8 and Fig.~9.
Consider as an example OGLE-TR-5, with its spectrum
shown in Fig. 4.  The peak corresponding to the ellipsoidal variability
at $ 2/\porb = 2.48 \invday $ is strong and certainly real.  The 
amplitude is $ a_{c2} = 7.5 \,\rm{mmag} $ 
in our Table 1, and 7.2 mmag in Drake's
Table 1.  However, the errors are 1.3 mmag and 0.4 mmag, respectively.  The
power spectrum presented in Fig. 4 clearly shows a high level of noise, which
is used for our error estimate.  

In the case of OGLE-TR-5 the ellipsoidal 
variability is highly significant with either of the two error estimates.
It is not so with OGLE-TR-40, which is listed by Drake as being ellipsoidal
variable at the $ 3.5 \sigma $ level, while our analysis puts it at just a
one $ \sigma $ level, i.e. nothing definite can be said about this case.

The star with the first planetary 
companion confirmed spectroscopically, OGLE-TR-56 (Konacki et al. 
2003), should not have a measurable value of $a_{c2}$, 
and reassuringly we do not detect a significant 
ellipsoidal variability (cf. Table 1).   The planetary disk covers
$ \sim 2 \times 10^{-4} $ of the sky as seen from the star, i.e. the 
reflection effect has to be small, $ a_{c1} < 0.1 ~ \rm{mmag} $.  The
measured value is $ a_{c1} = 1.08 \pm 0.46 ~ \rm{mmag} $, and presumably
it is not significant.

A thorough analysis of various systematic effects apparent in the 
photometry of tens of thousands of variable and non-variable stars measured 
in the OGLE fields is currently being done by Kruszewski \& Semeniuk
(2003).  Preliminary results indicate that various systematic errors 
may be reduced considerably.  This may allow a detection of smaller 
ellipsoidal effects than we could find, and may provide a
cleaner list of systems which are likely to have planetary companions.
At this time
spectroscopists may use our Table 1 to select stars for their planetary
search, eliminating objects with measurable ellipsoidal variability,
as it implies a large mass ratio, and most likely a red dwarf companion.

If the objective of this work was to identify stars with definite
ellipsoidal variability, we would select stars with their $ a_{c2} $ terms
positive at the several $ \sigma $ level.  But our objective is different: we
identify stars which are the best candidates to have planetary
companions, i.e. stars without ellipsoidal variability.  This can be done
in a statistical sense only.  The best candidates are those for which
the $ a_{c2} $ term is either negative or positive but small.  However,
even if we had perfect statistical information we could only assign a
probability that a given star does or does not have ellipsoidal variability.
This would always be in the sense that the smaller the $ a_{c2} $ term
relative to its error, the more likely the star is a planetary system.
Given limited access to big telescopes needed for spectroscopic follow-up
it is best to study the stars with the smallest (i.e., negative) 
$ a_{c2} $ terms first, and gradually move `up' in $a_{c2}/b_2$ from Table 1.

There are several developments which will improve the situation gradually.
First, OGLE continues its `planetary campaigns' and new candidate stars
will be published periodically.  Second, a highly improved analysis of
photometry is under way (Kruszewski \& Semeniuk 2003), which will reduce the
correlation of consecutive photometric data points considerably.  
This will allow a detection of smaller
amplitude ellipsoidal variations, and a much better rejection of stars
with non-planetary companions.  It will also be possible to identify
stars with even smaller depth of transits, extending the current list.

It is a pleasure to acknowledge many useful suggestions and discussions with 
A. Kruszewski, R. Lupton, S. Ruci\'nski and A. Udalski.  We thank the
referee for many useful suggestions and for identifying the star with
two names.  This research was supported
by the NASA grant NAG5-12212 and the NSF grant AST-0204908.


\begin{deluxetable}{lrrrrr}
\tablewidth{0pt}
\tablecaption{List of transiting objects}
\tablecomments{We present the $a_{c1}$ and $a_{c2}$ coefficients in the
format $b_i(a_{ci}/b_i \pm 1)$ to more easily identify the ratio of
the coefficients to their errors. 
*OGLE-TR-8 and OGLE-TR-29 are the same
object, but recorded as two separate events by OGLE. We treat them
separately throughout this paper.}
\tablehead{\colhead{Object name} & \colhead{$<I>$} & \colhead{$\porb$} &
\colhead{$a_{c1}$} & \colhead{$a_{c2}$} & \colhead{$N_{\rm{tr}}$} }
\startdata
\input{transit_table.tex}
\enddata
\end{deluxetable}

\begin{figure}
\includegraphics[width=\textwidth]{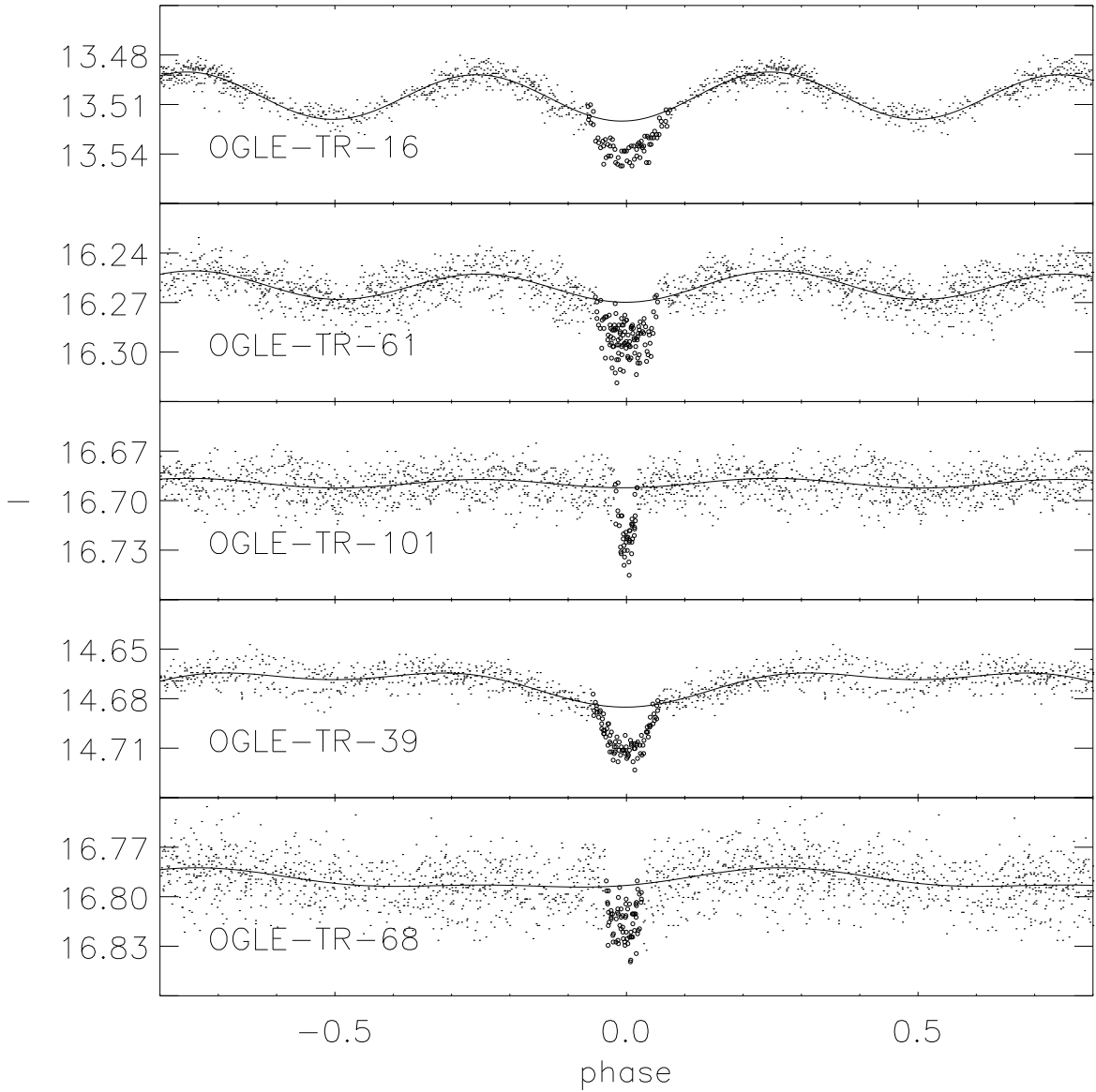}
\caption{Light curves of five transiting objects, folded with the transit
at zero phase, are shown as examples.  
Photometric data points within the transits, indicated
with circles, are not used in fitting.  The best-fit five-parameter
function (eq. 1) is also plotted.
Objects OGLE-TR-16 and TR-61 have ellipsoidal variability at
about the 8$\sigma$ level, while OGLE-TR-101 has ellipsoidal variability
at about a 4$\sigma$ level (cf. Table 1).  OGLE-TR-39 has a strong 
$a_{c1}$ term, indicative of surface heating (cf. Fig. 7 and 8), as
well as a strong ellipsoidal variability term $a_{c2}$.
OGLE-TR-68 has a strong $a_{s1}$ term (cf. Fig. 9), indicating a probable
spot activity.}
\end{figure}

\begin{figure}
\includegraphics[width=\textwidth]{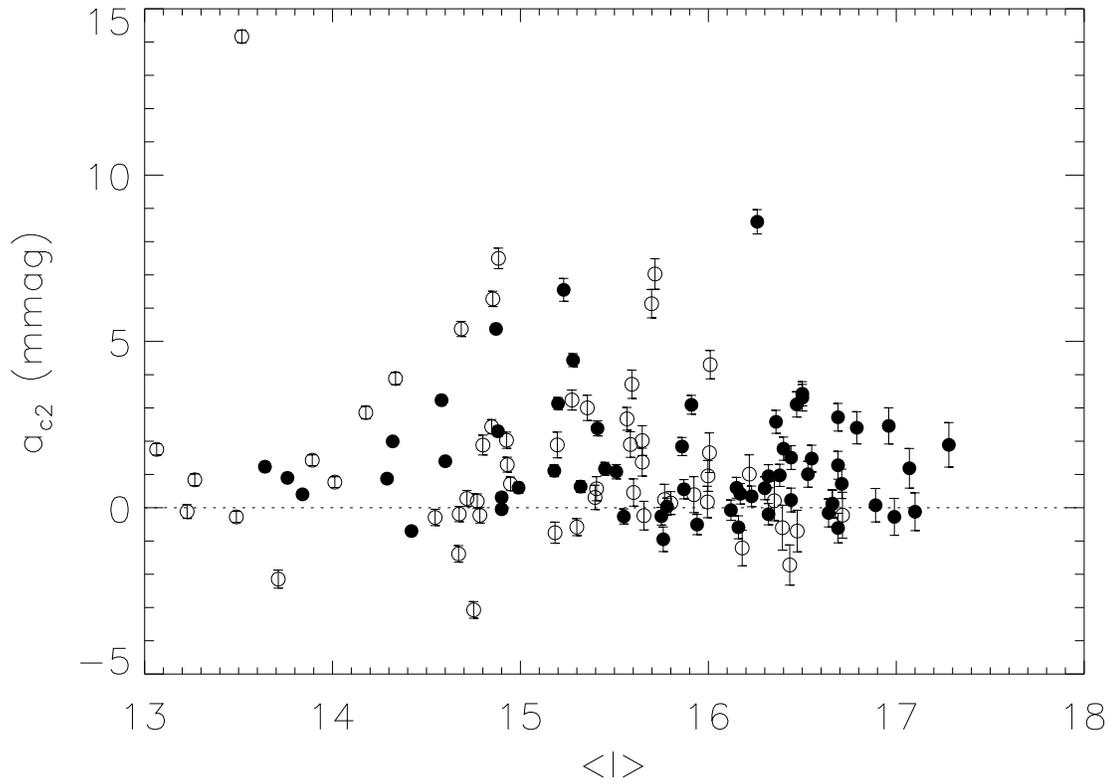} 
\caption{
The values of the parameter corresponding to ellipsoidal light variations
are shown as a function of average stellar magnitude $ <I> $.  Open and 
filled symbols refer to the data sets OGLE-TR-1 to TR-59 and TR-60 to TR-121, 
respectively.  Formal errors obtained
with the least squares fit are also shown.  Note a number of stars with
negative, i.e. not physical, values of $ a_{c2} $, which are many formal
standard deviations below zero.  This indicates that the errors are not
realistic.
}
\end{figure}

\begin{figure}
\includegraphics[width=\textwidth]{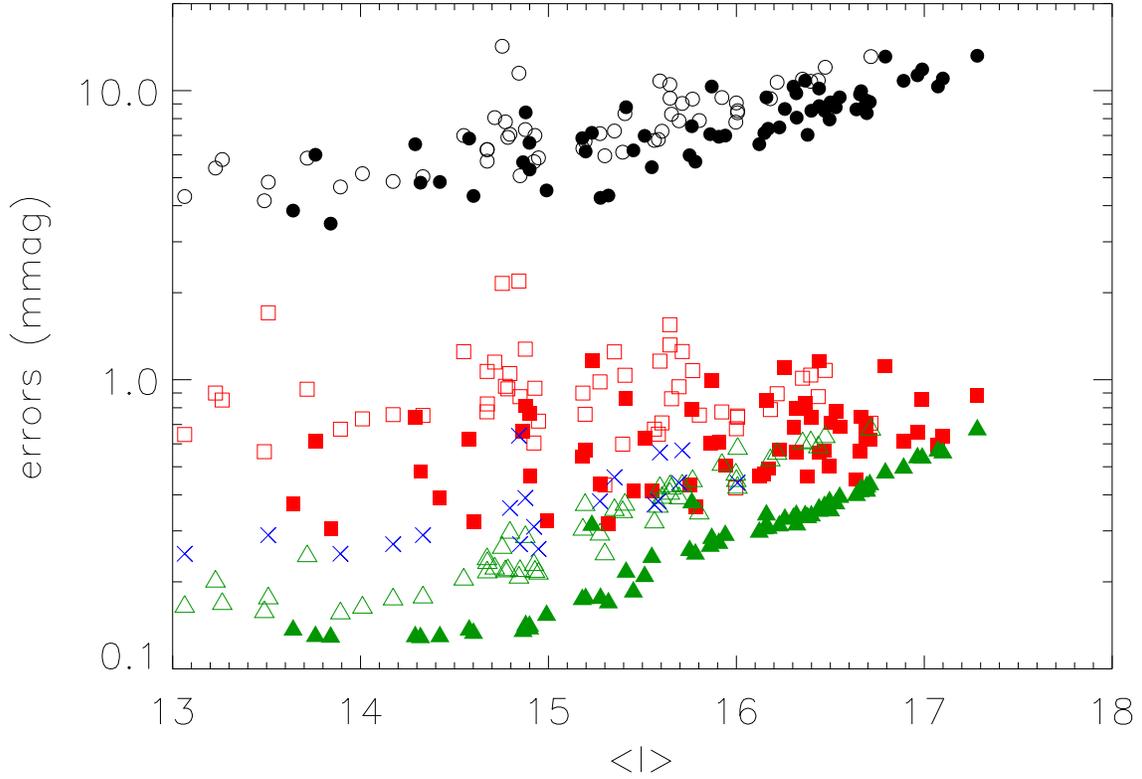} 
\caption{
Photometric errors are shown as a function of $ <I> $ magnitude.  Circles
refer to the errors of individual data points, defined as an rms deviation
from a 5 parameter fit to the light curves outside of the transits 
(cf. eq. 1).  The two outliers near $ <I>\ \approx 14.8 $, stars TR-24 and 
TR-58, have a long term variability at the level of 0.015 mag.   
Triangles are the average of the formal errors of the 5 parameter fit 
obtained with the least squares solution of eq. (1).  Squares are 
the errors of the $a_{c2}$ term determined with our spectral analysis - note
they are considerably larger than the formal errors, as the photometric data
have a strong time correlation.  Open and filled symbols refer to 
the OGLE-TR-1 to TR-59 and TR-60 to TR-121 data sets, respectively.
Errors listed by Drake (2003) in his Table 1
are shown with crosses for stars in the Galactic Center region.
}
\end{figure}

\begin{figure}
\includegraphics[width=\textwidth]{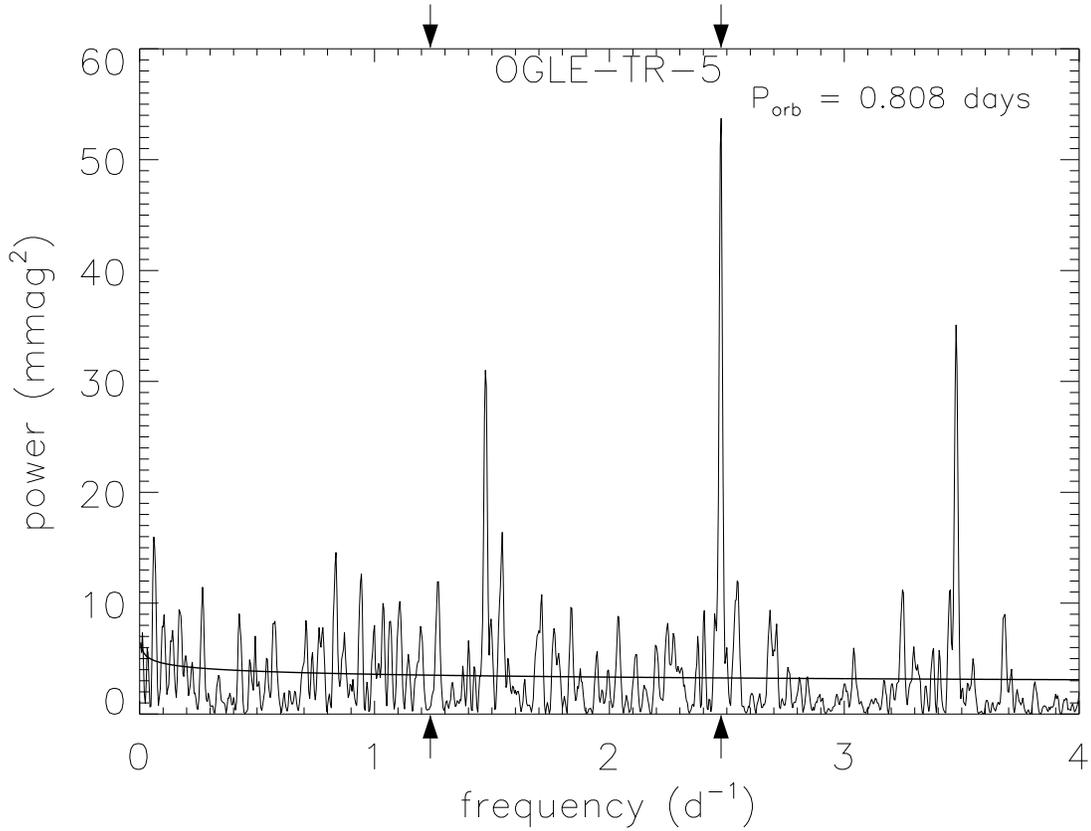} 
\caption{
The power spectrum is shown for OGLE-TR-5.  The thick solid 
line is the least squares power law fit to the spectrum.  The two arrows
correspond to the orbital frequency $ 1 / \porb = 1.237 \invday $,
and the frequency of expected ellipsoidal variability 
$ 2 / \porb = 2.475 \invday $.  Notice a very strong peak at
the `ellipsoidal' frequency, and the two aliases at $ 1.475 \invday $ and
at $ 3.475 \invday $.
}
\end{figure}

\begin{figure}
\includegraphics[width=\textwidth]{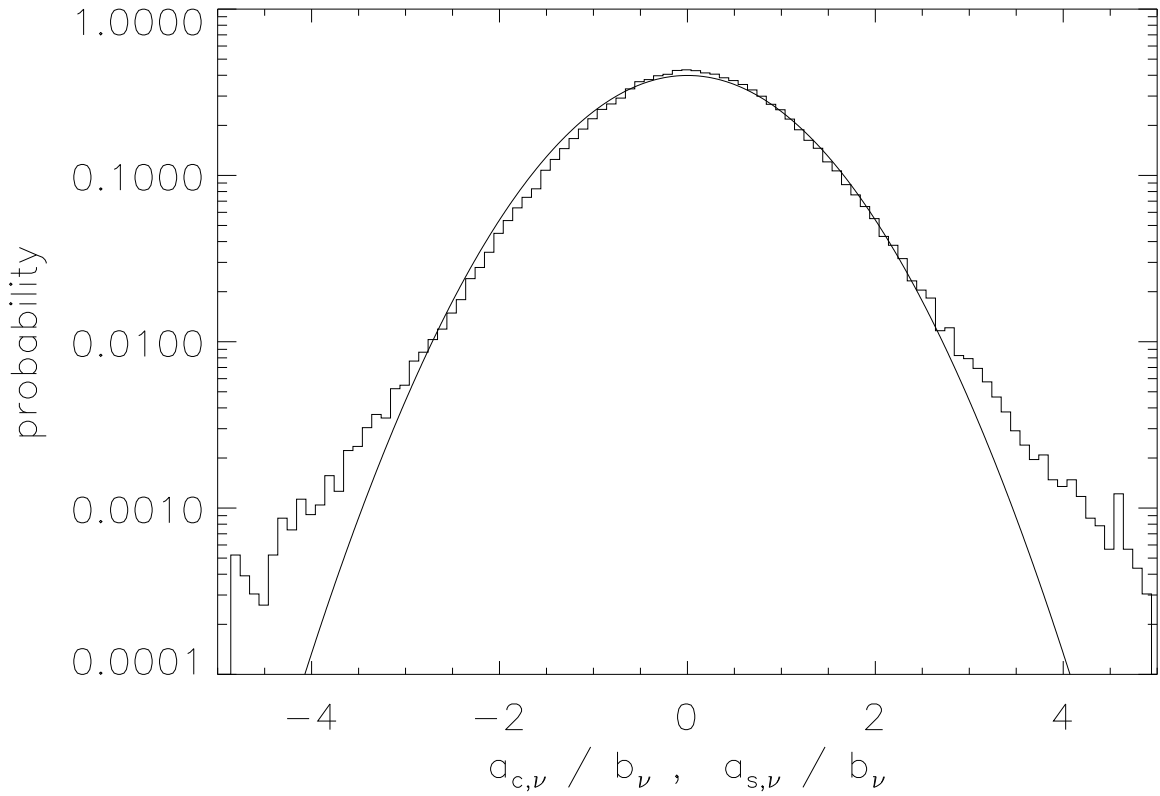} 
\caption{
The distribution of normalized amplitudes of sine and cosine terms for
all sampled frequencies and for all transit candidates except OGLE-TR-24
and TR-58.  The ordinate axis is logarithmic to bring out the tails
in the distribution.  For comparison
a Gaussian distribution with unit variance is also shown.
}
\end{figure}

\begin{figure}
\includegraphics[width=\textwidth]{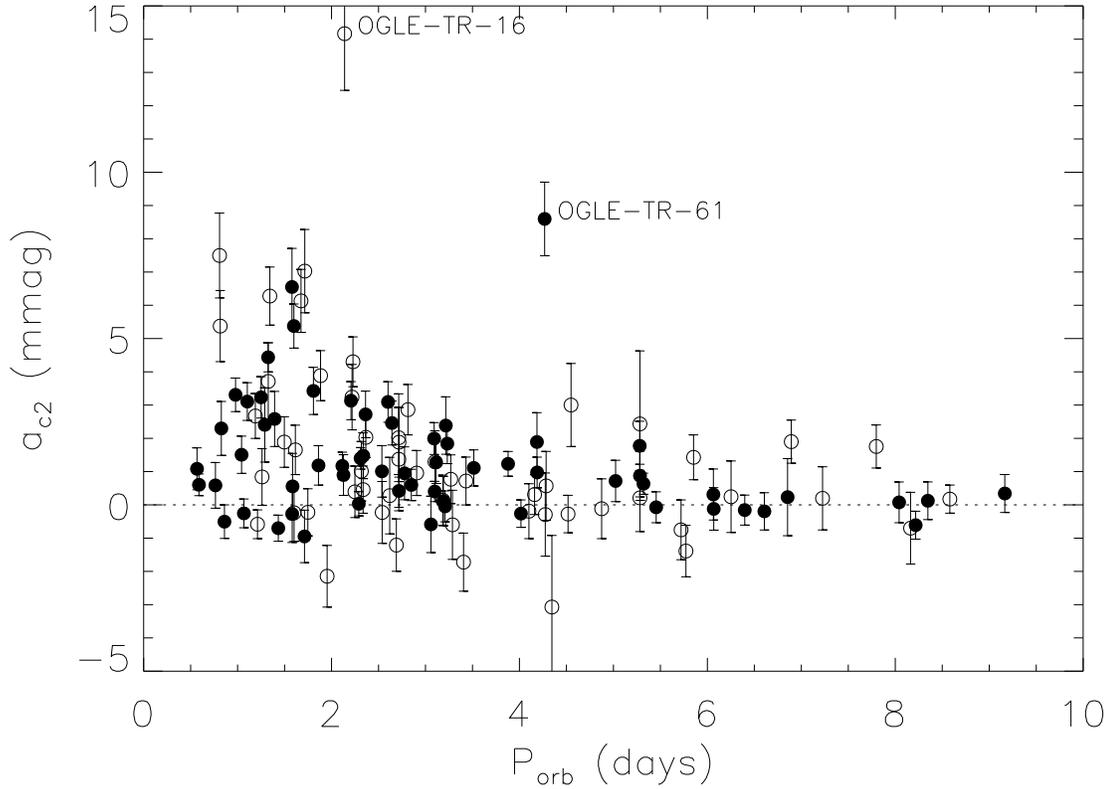} 
\caption{
The values of the parameter $ a_{c2} $, corresponding to ellipsoidal light
variations, are shown as a function of orbital period, $ \porb $.  Open
and filled symbols refer to the stars OGLE-TR-1 to TR-59, and TR-60 to TR-121, 
respectively.  The errors are based on the limited spectral analysis
presented in this paper.  None of the negative values are significant.
Up to $ \sim 50\% $ of all stars may have significant ellipsoidal
variability, indicating a massive, not planetary, companion.
}
\end{figure}

\begin{figure}
\includegraphics[width=\textwidth]{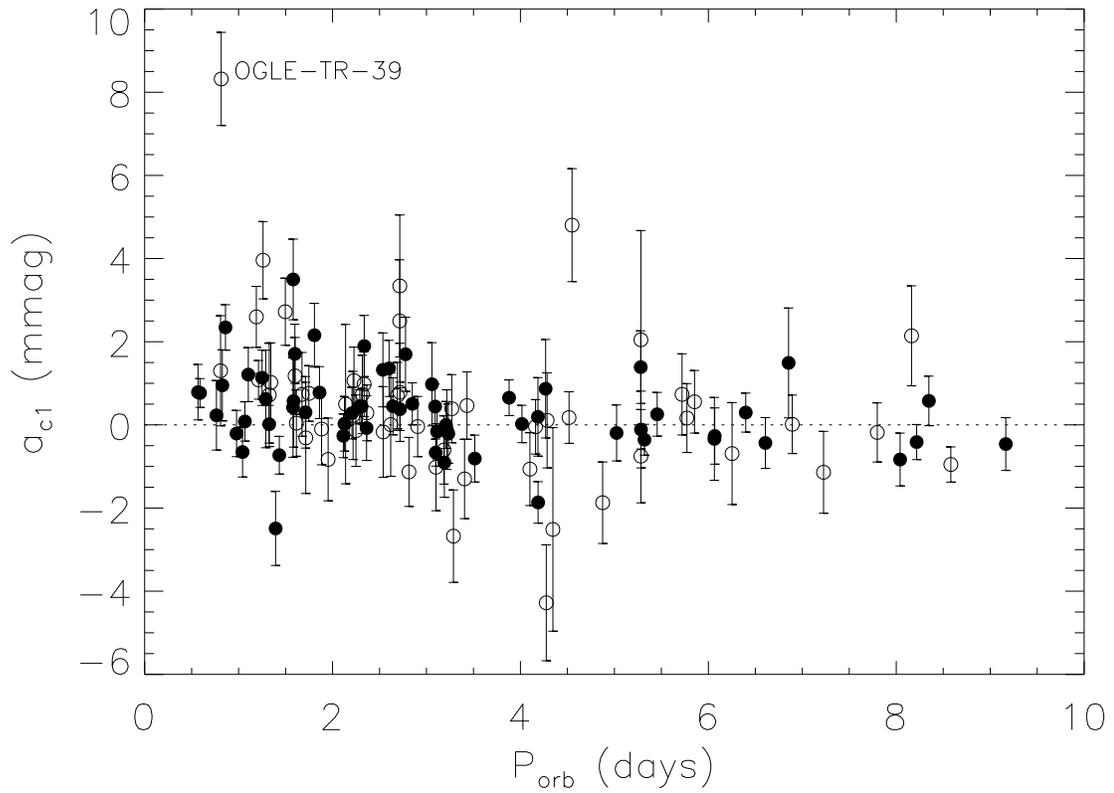} 
\caption{
The same as Fig. 6, but for the $ a_{c1} $ parameter, which is indicative
of possible `heating' (reflection) effects of the companion's hemisphere 
facing the primary.  Only OGLE-TR-39 shows the effect strongly, but there 
may be several other stars for which the effect is real.
}
\end{figure}

\begin{figure}
\includegraphics[width=\textwidth]{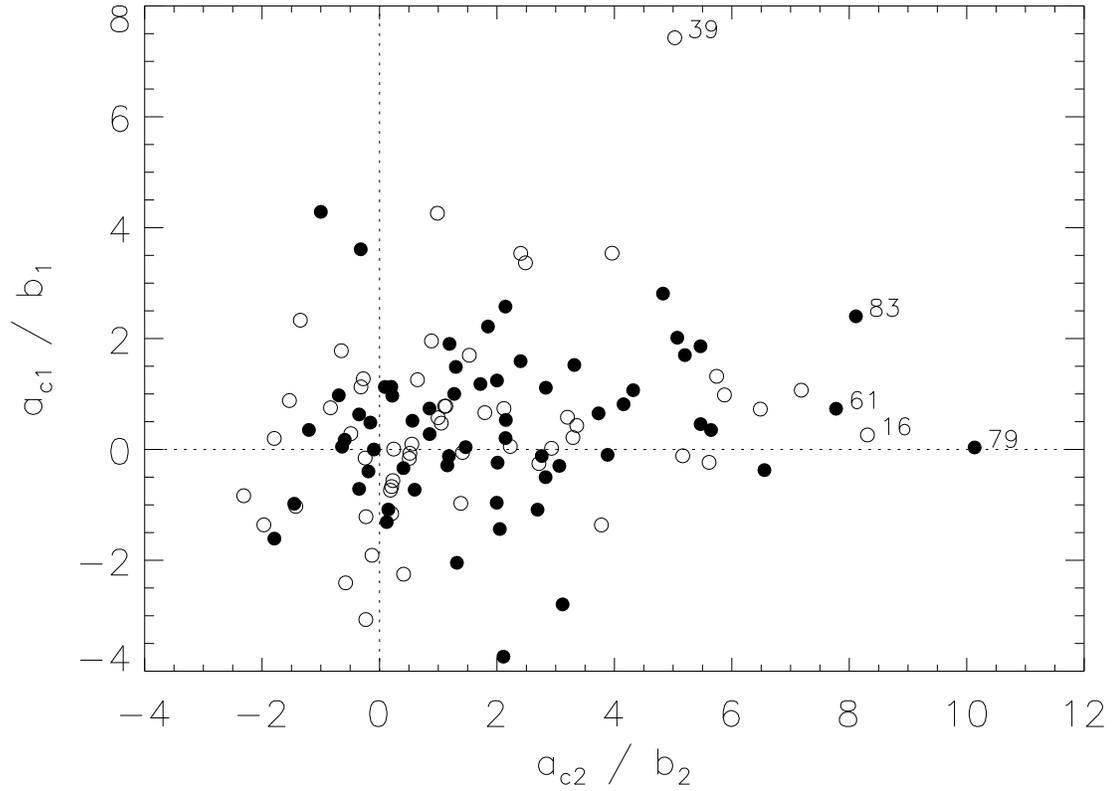} 
\caption{
A relation between the ellipsoidal variability parameter $ a_{c2} $, and the
heating variability parameter $ a_{c1} $ (cf. eq. 1), both
normalized by their errors.  The stars with the
strongest effects are labeled with their OGLE names.  The negative
values of either parameter are consistent with them being due to errors.
}
\end{figure}

\begin{figure}[t]
\includegraphics[width=\textwidth]{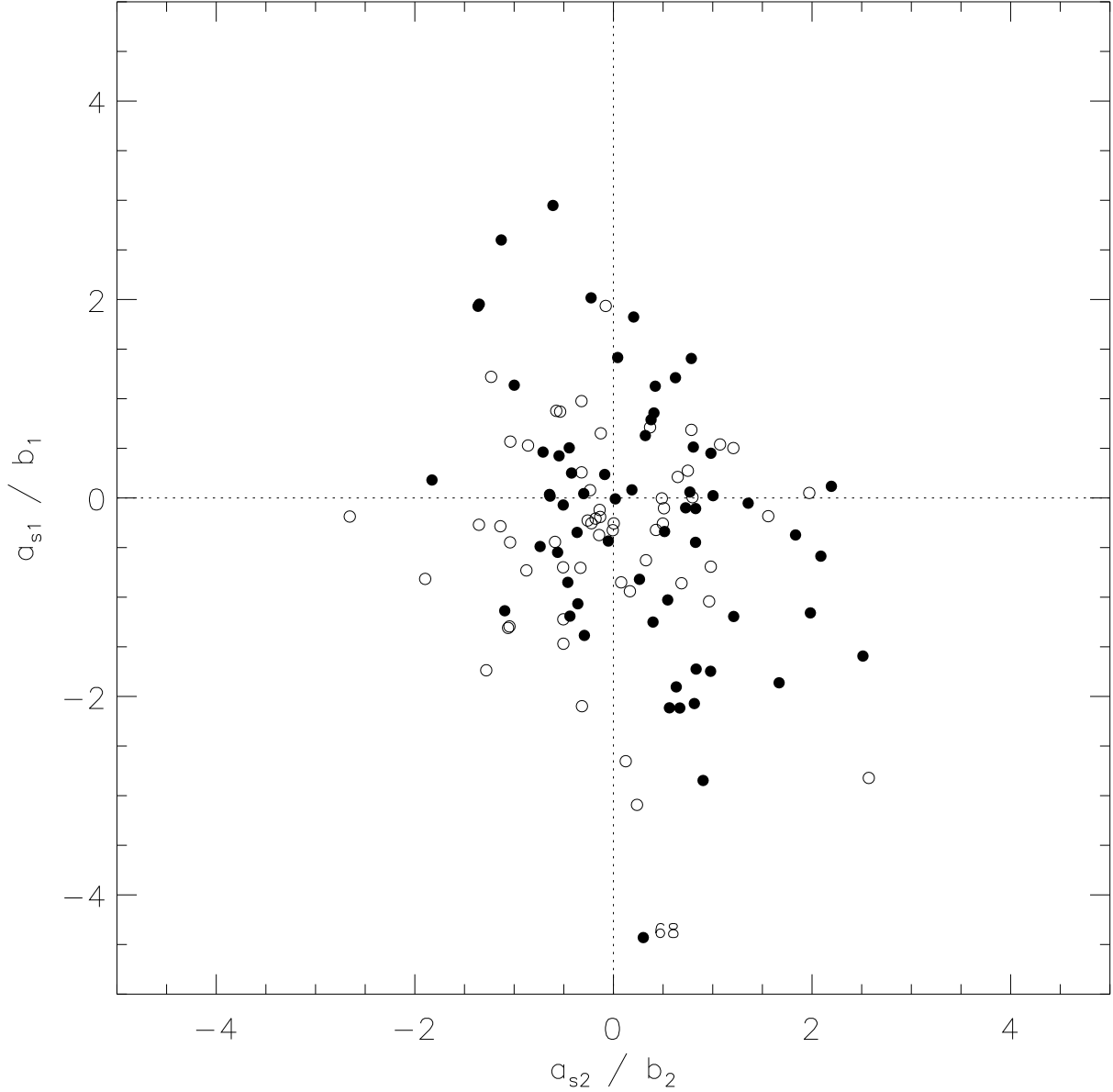} 
\caption{
A relation between the coefficients of the sine terms in eq. (1),
normalized by their errors.
The ratio of the number of stars having $a_{s1}$ within one $\sigma$ 
of zero to the number outside of one $\sigma$ is approximately 2/1,
and for $a_{s2}$ the ratio is approximately 3/1.  We expect these ratios
to be 2/1 if the true values are zero and the distributions are Gaussian.
We conclude that our error estimate is reasonable, and we have no 
coefficients $ a_{s1} $ or $ a_{s2} $ that are measurably
non-zero, with the possible exception of OGLE-TR-68, the filled
circle with $ a_{s1} / b_1 = -4.4 $.  Visual inspection of its light curve
clearly shows that the system is approximately 15 mmag brighter at phase
0.25 than it is at phase 0.75.
}
\end{figure}

\end{document}

%% file: transit_table.tex
  OGLE-TR-1  &   15.655 &    1.601 &    $0.92~(+1.27 \pm 1)$ &    $0.86~(-0.28 \pm 1)$ &   4 \\
  OGLE-TR-2  &   14.173 &    2.813 &    $0.83~(-1.36 \pm 1)$ &    $0.76~(+3.78 \pm 1)$ &   5 \\
  OGLE-TR-3  &   15.564 &    1.189 &    $0.73~(+3.54 \pm 1)$ &    $0.67~(+3.96 \pm 1)$ &  14 \\
  OGLE-TR-4  &   14.714 &    2.619 &    $1.24~(+0.01 \pm 1)$ &    $1.15~(+0.24 \pm 1)$ &   5 \\
  OGLE-TR-5  &   14.877 &    0.808 &    $1.32~(+0.98 \pm 1)$ &    $1.28~(+5.87 \pm 1)$ &   8 \\
  OGLE-TR-6  &   15.351 &    4.549 &    $1.36~(+3.54 \pm 1)$ &    $1.25~(+2.40 \pm 1)$ &   4 \\
  OGLE-TR-7  &   14.795 &    2.718 &    $1.18~(+0.66 \pm 1)$ &    $1.05~(+1.79 \pm 1)$ &   5 \\
  OGLE-TR-8* &   15.647 &    2.715 &    $1.47~(+1.70 \pm 1)$ &    $1.32~(+1.53 \pm 1)$ &   4 \\
  OGLE-TR-9  &   14.010 &    3.269 &    $0.82~(+0.47 \pm 1)$ &    $0.73~(+1.05 \pm 1)$ &   4 \\
 OGLE-TR-10  &   14.928 &    3.101 &    $1.05~(-0.97 \pm 1)$ &    $0.93~(+1.38 \pm 1)$ &   7 \\
 OGLE-TR-11  &   16.007 &    1.615 &    $0.80~(+0.05 \pm 1)$ &    $0.74~(+2.23 \pm 1)$ &   7 \\
 OGLE-TR-12  &   14.673 &    5.772 &    $0.83~(+0.20 \pm 1)$ &    $0.77~(-1.79 \pm 1)$ &   2 \\
 OGLE-TR-13  &   13.893 &    5.853 &    $0.75~(+0.74 \pm 1)$ &    $0.67~(+2.12 \pm 1)$ &   2 \\
 OGLE-TR-14  &   13.064 &    7.798 &    $0.71~(-0.25 \pm 1)$ &    $0.65~(+2.71 \pm 1)$ &   3 \\
 OGLE-TR-15  &   13.228 &    4.875 &    $0.98~(-1.91 \pm 1)$ &    $0.90~(-0.13 \pm 1)$ &   4 \\
 OGLE-TR-16  &   13.509 &    2.139 &    $1.92~(+0.26 \pm 1)$ &    $1.70~(+8.31 \pm 1)$ &   4 \\
 OGLE-TR-17  &   16.217 &    2.317 &    $0.94~(+0.78 \pm 1)$ &    $0.89~(+1.13 \pm 1)$ &   4 \\
 OGLE-TR-18  &   16.006 &    2.228 &    $0.81~(+1.32 \pm 1)$ &    $0.75~(+5.74 \pm 1)$ &   6 \\
 OGLE-TR-19  &   16.352 &    5.282 &    $1.12~(-0.67 \pm 1)$ &    $1.01~(+0.21 \pm 1)$ &   2 \\
 OGLE-TR-20  &   15.407 &    4.284 &    $1.14~(+0.10 \pm 1)$ &    $1.04~(+0.55 \pm 1)$ &   3 \\
 OGLE-TR-21  &   15.585 &    6.893 &    $0.70~(+0.02 \pm 1)$ &    $0.65~(+2.93 \pm 1)$ &   3 \\
 OGLE-TR-22  &   14.549 &    4.275 &    $1.39~(-3.07 \pm 1)$ &    $1.25~(-0.23 \pm 1)$ &   3 \\
 OGLE-TR-23  &   16.396 &    3.287 &    $1.11~(-2.41 \pm 1)$ &    $1.04~(-0.58 \pm 1)$ &   2 \\
 OGLE-TR-24  &   14.843 &    5.282 &    $2.63~(+0.78 \pm 1)$ &    $2.20~(+1.11 \pm 1)$ &   2 \\
 OGLE-TR-25  &   15.274 &    2.218 &    $1.06~(+0.21 \pm 1)$ &    $0.98~(+3.29 \pm 1)$ &   5 \\
 OGLE-TR-26  &   14.784 &    2.539 &    $1.09~(-0.16 \pm 1)$ &    $0.93~(-0.25 \pm 1)$ &   4 \\
 OGLE-TR-27  &   15.712 &    1.715 &    $1.34~(-0.23 \pm 1)$ &    $1.25~(+5.61 \pm 1)$ &   6 \\
 OGLE-TR-28  &   16.436 &    3.405 &    $0.96~(-1.36 \pm 1)$ &    $0.87~(-1.97 \pm 1)$ &   3 \\
 OGLE-TR-29* &   15.646 &    2.716 &    $1.71~(+1.96 \pm 1)$ &    $1.55~(+0.88 \pm 1)$ &   4 \\
 OGLE-TR-30  &   14.923 &    2.365 &    $0.67~(+0.43 \pm 1)$ &    $0.60~(+3.36 \pm 1)$ &   6 \\
 OGLE-TR-31  &   14.332 &    1.883 &    $0.86~(-0.12 \pm 1)$ &    $0.75~(+5.16 \pm 1)$ &   7 \\
 OGLE-TR-32  &   14.849 &    1.343 &    $0.95~(+1.07 \pm 1)$ &    $0.87~(+7.18 \pm 1)$ &   7 \\
 OGLE-TR-33  &   13.716 &    1.953 &    $0.99~(-0.84 \pm 1)$ &    $0.93~(-2.31 \pm 1)$ &   4 \\
 OGLE-TR-34  &   15.997 &    8.581 &    $0.42~(-2.25 \pm 1)$ &    $0.42~(+0.41 \pm 1)$ &   3 \\
 OGLE-TR-35  &   13.264 &    1.260 &    $0.93~(+4.26 \pm 1)$ &    $0.85~(+0.99 \pm 1)$ &   7 \\
 OGLE-TR-36  &   15.767 &    6.252 &    $1.23~(-0.56 \pm 1)$ &    $1.08~(+0.23 \pm 1)$ &   2 \\
 OGLE-TR-37  &   15.183 &    5.720 &    $0.98~(+0.75 \pm 1)$ &    $0.90~(-0.84 \pm 1)$ &   2 \\
 OGLE-TR-38  &   14.675 &    4.101 &    $0.88~(-1.21 \pm 1)$ &    $0.82~(-0.23 \pm 1)$ &   3 \\
 OGLE-TR-39  &   14.674 &    0.815 &    $1.12~(+7.43 \pm 1)$ &    $1.07~(+5.03 \pm 1)$ &  11 \\
 OGLE-TR-40  &   14.947 &    3.431 &    $0.81~(+0.57 \pm 1)$ &    $0.72~(+1.00 \pm 1)$ &   5 \\
 OGLE-TR-41  &   13.488 &    4.517 &    $0.62~(+0.28 \pm 1)$ &    $0.56~(-0.49 \pm 1)$ &   2 \\
 OGLE-TR-42  &   15.395 &    4.161 &    $0.65~(-0.07 \pm 1)$ &    $0.60~(+0.52 \pm 1)$ &   4 \\
 OGLE-TR-47  &   15.604 &    2.336 &    $0.78~(+1.26 \pm 1)$ &    $0.71~(+0.65 \pm 1)$ &   7 \\
 OGLE-TR-48  &   14.771 &    7.226 &    $0.99~(-1.16 \pm 1)$ &    $0.95~(+0.21 \pm 1)$ &   2 \\
 OGLE-TR-49  &   16.181 &    2.690 &    $0.80~(+0.88 \pm 1)$ &    $0.79~(-1.53 \pm 1)$ &   2 \\
 OGLE-TR-50  &   15.923 &    2.249 &    $0.86~(-0.16 \pm 1)$ &    $0.77~(+0.51 \pm 1)$ &   3 \\
 OGLE-TR-51  &   16.716 &    1.748 &    $0.67~(+1.13 \pm 1)$ &    $0.71~(-0.32 \pm 1)$ &   5 \\
 OGLE-TR-52  &   15.593 &    1.326 &    $1.24~(+0.58 \pm 1)$ &    $1.16~(+3.20 \pm 1)$ &   8 \\
 OGLE-TR-53  &   16.000 &    2.906 &    $0.72~(-0.06 \pm 1)$ &    $0.68~(+1.41 \pm 1)$ &   4 \\
 OGLE-TR-54  &   16.473 &    8.163 &    $1.20~(+1.78 \pm 1)$ &    $1.08~(-0.65 \pm 1)$ &   3 \\
 OGLE-TR-55  &   15.803 &    3.185 &    $0.82~(-0.74 \pm 1)$ &    $0.75~(+0.19 \pm 1)$ &   6 \\
 OGLE-TR-56  &   15.300 &    1.212 &    $0.46~(+2.33 \pm 1)$ &    $0.43~(-1.35 \pm 1)$ &  11 \\
 OGLE-TR-57  &   15.695 &    1.675 &    $1.01~(+0.73 \pm 1)$ &    $0.95~(+6.49 \pm 1)$ &   5 \\
 OGLE-TR-58  &   14.754 &    4.345 &    $2.45~(-1.03 \pm 1)$ &    $2.15~(-1.43 \pm 1)$ &   3 \\
 OGLE-TR-59  &   15.195 &    1.497 &    $0.81~(+3.37 \pm 1)$ &    $0.76~(+2.49 \pm 1)$ &   9 \\
 OGLE-TR-60  &   14.601 &    2.309 &    $0.42~(+1.07 \pm 1)$ &    $0.32~(+4.32 \pm 1)$ &  11 \\
 OGLE-TR-61  &   16.258 &    4.268 &    $1.18~(+0.73 \pm 1)$ &    $1.11~(+7.77 \pm 1)$ &   8 \\
 OGLE-TR-62  &   15.907 &    2.601 &    $0.67~(+2.02 \pm 1)$ &    $0.61~(+5.07 \pm 1)$ &  10 \\
 OGLE-TR-63  &   15.751 &    1.067 &    $0.47~(+0.17 \pm 1)$ &    $0.43~(-0.59 \pm 1)$ &  12 \\
 OGLE-TR-64  &   16.169 &    2.717 &    $0.51~(+0.74 \pm 1)$ &    $0.49~(+0.85 \pm 1)$ &   7 \\
 OGLE-TR-65  &   15.941 &    0.860 &    $0.55~(+4.29 \pm 1)$ &    $0.51~(-1.00 \pm 1)$ &  18 \\
 OGLE-TR-66  &   15.180 &    3.514 &    $0.56~(-1.44 \pm 1)$ &    $0.54~(+2.05 \pm 1)$ &   6 \\
 OGLE-TR-67  &   16.399 &    5.280 &    $0.87~(+1.59 \pm 1)$ &    $0.74~(+2.40 \pm 1)$ &   5 \\
 OGLE-TR-68  &   16.793 &    1.289 &    $1.17~(+0.53 \pm 1)$ &    $1.12~(+2.15 \pm 1)$ &  12 \\
 OGLE-TR-69  &   16.550 &    2.337 &    $0.74~(+2.58 \pm 1)$ &    $0.69~(+2.14 \pm 1)$ &   5 \\
 OGLE-TR-70  &   16.890 &    8.041 &    $0.64~(-1.31 \pm 1)$ &    $0.61~(+0.12 \pm 1)$ &   4 \\
 OGLE-TR-71  &   16.379 &    4.188 &    $0.50~(-3.74 \pm 1)$ &    $0.46~(+2.11 \pm 1)$ &   5 \\
 OGLE-TR-72  &   16.440 &    6.854 &    $1.32~(+1.13 \pm 1)$ &    $1.16~(+0.20 \pm 1)$ &   4 \\
 OGLE-TR-73  &   16.989 &    1.581 &    $0.97~(+3.61 \pm 1)$ &    $0.86~(-0.32 \pm 1)$ &   9 \\
 OGLE-TR-74  &   15.869 &    1.585 &    $1.11~(+0.52 \pm 1)$ &    $0.99~(+0.56 \pm 1)$ &  11 \\
 OGLE-TR-75  &   16.964 &    2.643 &    $0.69~(+0.65 \pm 1)$ &    $0.66~(+3.73 \pm 1)$ &   8 \\
 OGLE-TR-76  &   13.760 &    2.127 &    $0.66~(+0.04 \pm 1)$ &    $0.61~(+1.47 \pm 1)$ &   6 \\
 OGLE-TR-77  &   16.122 &    5.455 &    $0.53~(+0.48 \pm 1)$ &    $0.47~(-0.16 \pm 1)$ &   4 \\
 OGLE-TR-78  &   15.319 &    5.320 &    $0.37~(-0.96 \pm 1)$ &    $0.32~(+2.00 \pm 1)$ &   4 \\
 OGLE-TR-79  &   15.277 &    1.325 &    $0.45~(+0.04 \pm 1)$ &   $0.44~(+10.13 \pm 1)$ &  13 \\
 OGLE-TR-80  &   16.501 &    1.807 &    $0.77~(+2.81 \pm 1)$ &    $0.71~(+4.83 \pm 1)$ &  12 \\
 OGLE-TR-81  &   15.413 &    3.216 &    $0.96~(-0.12 \pm 1)$ &    $0.86~(+2.76 \pm 1)$ &   6 \\
 OGLE-TR-82  &   16.304 &    0.764 &    $0.84~(+0.28 \pm 1)$ &    $0.69~(+0.85 \pm 1)$ &  22 \\
 OGLE-TR-83  &   14.865 &    1.599 &    $0.71~(+2.40 \pm 1)$ &    $0.66~(+8.11 \pm 1)$ &  12 \\
 OGLE-TR-84  &   16.692 &    3.113 &    $0.68~(-0.24 \pm 1)$ &    $0.64~(+2.01 \pm 1)$ &   6 \\
 OGLE-TR-85  &   15.452 &    2.115 &    $0.53~(-0.50 \pm 1)$ &    $0.41~(+2.83 \pm 1)$ &  12 \\
 OGLE-TR-86  &   16.319 &    2.777 &    $0.89~(+1.90 \pm 1)$ &    $0.80~(+1.19 \pm 1)$ &   7 \\
 OGLE-TR-87  &   16.321 &    6.607 &    $0.61~(-0.71 \pm 1)$ &    $0.56~(-0.35 \pm 1)$ &   3 \\
 OGLE-TR-88  &   14.578 &    1.250 &    $0.67~(+1.70 \pm 1)$ &    $0.62~(+5.20 \pm 1)$ &  15 \\
 OGLE-TR-89  &   15.782 &    2.290 &    $0.40~(+1.13 \pm 1)$ &    $0.36~(+0.09 \pm 1)$ &   5 \\
 OGLE-TR-90  &   16.441 &    1.042 &    $0.60~(-1.09 \pm 1)$ &    $0.56~(+2.69 \pm 1)$ &  15 \\
 OGLE-TR-91  &   15.231 &    1.579 &    $1.20~(+0.35 \pm 1)$ &    $1.16~(+5.65 \pm 1)$ &   9 \\
 OGLE-TR-92  &   16.496 &    0.978 &    $0.56~(-0.37 \pm 1)$ &    $0.50~(+6.55 \pm 1)$ &  20 \\
 OGLE-TR-93  &   15.198 &    2.207 &    $0.62~(+0.46 \pm 1)$ &    $0.57~(+5.47 \pm 1)$ &  12 \\
 OGLE-TR-94  &   14.319 &    3.092 &    $0.54~(+0.81 \pm 1)$ &    $0.48~(+4.16 \pm 1)$ &   6 \\
 OGLE-TR-95  &   16.366 &    1.394 &    $0.89~(-2.80 \pm 1)$ &    $0.83~(+3.12 \pm 1)$ &  14 \\
 OGLE-TR-96  &   14.900 &    3.208 &    $0.56~(-0.00 \pm 1)$ &    $0.47~(-0.10 \pm 1)$ &   6 \\
 OGLE-TR-97  &   15.512 &    0.568 &    $0.67~(+1.18 \pm 1)$ &    $0.63~(+1.72 \pm 1)$ &  25 \\
 OGLE-TR-98  &   16.639 &    6.398 &    $0.47~(+0.63 \pm 1)$ &    $0.45~(-0.35 \pm 1)$ &   5 \\
 OGLE-TR-99  &   16.469 &    1.103 &    $0.65~(+1.86 \pm 1)$ &    $0.57~(+5.47 \pm 1)$ &  16 \\
OGLE-TR-100  &   14.879 &    0.827 &    $0.85~(+1.11 \pm 1)$ &    $0.81~(+2.83 \pm 1)$ &  20 \\
OGLE-TR-101  &   16.689 &    2.362 &    $0.78~(-0.10 \pm 1)$ &    $0.70~(+3.88 \pm 1)$ &   8 \\
OGLE-TR-102  &   13.841 &    3.098 &    $0.33~(-2.05 \pm 1)$ &    $0.30~(+1.32 \pm 1)$ &   5 \\
OGLE-TR-103  &   16.694 &    8.217 &    $0.42~(-0.98 \pm 1)$ &    $0.42~(-1.45 \pm 1)$ &   4 \\
OGLE-TR-104  &   17.099 &    6.068 &    $0.68~(-0.40 \pm 1)$ &    $0.64~(-0.19 \pm 1)$ &   2 \\
OGLE-TR-105  &   16.160 &    3.058 &    $1.00~(+0.98 \pm 1)$ &    $0.85~(-0.69 \pm 1)$ &   3 \\
OGLE-TR-106  &   16.529 &    2.536 &    $0.89~(+1.49 \pm 1)$ &    $0.77~(+1.30 \pm 1)$ &   6 \\
OGLE-TR-107  &   16.664 &    3.190 &    $0.84~(-1.08 \pm 1)$ &    $0.74~(+0.15 \pm 1)$ &   7 \\
OGLE-TR-108  &   17.282 &    4.186 &    $0.94~(+0.20 \pm 1)$ &    $0.88~(+2.15 \pm 1)$ &   3 \\
OGLE-TR-109  &   14.990 &    0.589 &    $0.35~(+2.22 \pm 1)$ &    $0.33~(+1.85 \pm 1)$ &  24 \\
OGLE-TR-110  &   16.149 &    2.849 &    $0.51~(+1.00 \pm 1)$ &    $0.47~(+1.27 \pm 1)$ &   6 \\
OGLE-TR-111  &   15.550 &    4.016 &    $0.45~(+0.05 \pm 1)$ &    $0.41~(-0.64 \pm 1)$ &   9 \\
OGLE-TR-112  &   13.641 &    3.879 &    $0.43~(+1.52 \pm 1)$ &    $0.37~(+3.32 \pm 1)$ &   8 \\
OGLE-TR-113  &   14.422 &    1.433 &    $0.45~(-1.61 \pm 1)$ &    $0.39~(-1.79 \pm 1)$ &  10 \\
OGLE-TR-114  &   15.763 &    1.712 &    $0.86~(+0.35 \pm 1)$ &    $0.79~(-1.20 \pm 1)$ &   5 \\
OGLE-TR-115  &   16.658 &    8.347 &    $0.60~(+0.97 \pm 1)$ &    $0.57~(+0.22 \pm 1)$ &   3 \\
OGLE-TR-116  &   14.899 &    6.064 &    $1.00~(-0.34 \pm 1)$ &    $0.77~(+0.41 \pm 1)$ &   5 \\
OGLE-TR-117  &   16.710 &    5.023 &    $0.67~(-0.29 \pm 1)$ &    $0.62~(+1.15 \pm 1)$ &   5 \\
OGLE-TR-118  &   17.073 &    1.861 &    $0.63~(+1.24 \pm 1)$ &    $0.59~(+2.00 \pm 1)$ &   7 \\
OGLE-TR-119  &   14.291 &    5.283 &    $0.92~(-0.12 \pm 1)$ &    $0.74~(+1.18 \pm 1)$ &   7 \\
OGLE-TR-120  &   16.229 &    9.166 &    $0.63~(-0.73 \pm 1)$ &    $0.57~(+0.60 \pm 1)$ &   4 \\
OGLE-TR-121  &   15.861 &    3.232 &    $0.71~(-0.30 \pm 1)$ &    $0.60~(+3.06 \pm 1)$ &   6 \\